# CO$_2$ Laser-Induced Growth of Epitaxial Graphene on 6H-SiC(0001)

Spyros N. Yannopoulos,[1] Angeliki Siokou,[2] Nektarios K. Nasikas, Vassilios Dracopoulos, Fotini Ravani and George N. Papatheodorou

*Foundation for Research and Technology – Hellas, Institute of Chemical Engineering and High-Temperature Chemical Processes, (FORTH/ICE–HT), P.O. Box 1414, GR-26504, Rio-Patras, Greece*

[1]sny@iceht.forth.gr; [2]siokou@iceht.forth.gr

## Abstract

The thermal decomposition of SiC surface provides, perhaps, the most promising method for the epitaxial growth of graphene on a material useful in the electronics platform. Currently, efforts are focused on a reliable method for the growth of large-area, low-strain epitaxial graphene that is still lacking. We report here a novel method for the fast, single-step epitaxial growth of large-area homogeneous graphene film on the surface of SiC(0001) using an infrared CO$_2$ laser (10.6 μm) as the heating source. Apart from enabling extreme heating and cooling rates, which can control the stacking order of epitaxial graphene, this method is cost-effective in that it does not necessitate SiC pre-treatment and/or high vacuum, it operates at low temperature and proceeds in the second time scale, thus providing a green solution to EG fabrication and a means to engineering graphene patterns on SiC by focused laser beams. Uniform, low–strain graphene film is demonstrated by scanning electron microscopy and x-ray photoelectron, secondary ion mass, and Raman spectroscopies. Scalability to industrial level of the method described here appears to be realistic, in view of the high rate of CO$_2$-laser induced graphene growth and the lack of strict sample–environment conditions.

## 1. Introduction

Although graphene has been known for long time,[1,2] the isolation of free-standing monolayers[3] triggered extensive investigations over the last few years exploring its outstanding electronic, mechanical and thermal properties.[3,4] Three basic methods have been employed for graphene production, as reviewed recently.[2] Chemical vapor deposition (CVD) of hydrocarbons and epitaxial growth on transition metals' surfaces such as Ni or Cu has been successfully used for large area graphene growth.[5] Mechanical cleavage of graphene (MCG) from highly oriented pyrolitic graphite (HOPG) produces high quality, small size, graphene monolayers (ML),[3] which, although being useful for studying fundamental properties of two-dimensional crystals, their potential for electronics applications is still unclear. Perhaps, the most promising method for large area, high quality graphene production is the thermal decomposition of SiC due to Si sublimation, which leads to the growth of or epitaxial graphene (EG);[6] a method known since seventies.[7] The main advantage of this method is that graphene grows directly on a material useful in the electronics platform. The epitaxial growth method has itself several variants, it can proceed under ultrahigh vacuum (UHV) at ~1200 °C,[1] in the presence of inert gasses at ambient pressure or low vacuum and high temperature (~1550 °C),[8] and under other confinement controlled sublimation conditions of SiC.[1] Recently, EG was grown on SiC by molecular beam epitaxy.[9] Graphene grows on both the Si-terminated (0001) and the C-terminated ($000\bar{1}$) face (faster growth rate) of the SiC wafer, albeit with different electronic properties. Finally, other methods employ exfoliation of graphite oxide to form graphene oxide which can be reduced to become electrically active.[10]

The above methods for EG growth employ high temperature induction furnaces where the whole SiC wafer area is exposed to heat. Although EG produced in this way can be patterned using nanolithography methods[4,11,12] in order to be compatible with semiconductors technology, more convenient methods that make use of *in situ* patterning of graphene would be more valuable for the scale-up of the EG growth on an industrial scale. Along this direction, few studies have pointed out the possibility for simultaneous synthesis and patterning of graphene using laser beams.[13,14] In a recent investigation, EG was grown on the Si-face of SiC by the UV laser line (248 nm) of an KF excimer pulsed laser.[13] The surface-induced laser decomposition of the Si surface led to spatially controlled and scalable EG synthesis. The substrate was held at room temperature and the operation was carried in a vacuum chamber ($10^{-6}$ Torr). Using the visible (532 nm) line of a solid state laser to increase local heating, rapid printing of graphene patterns on nickel foil were achieved by means of CVD in a $CH_4$ / $H_2$ environment.[14] The substrate was held at room temperature and the graphene growth rate was estimated to be orders of magnitude faster than that of conventional CVD. Finally, in the pre-graphene era, it was observed that irradiation of polycrystalline SiC with a pulsed near-infrared Nd:YAG laser (1064 nm) yields "crystallized" graphite on the SiC surface.[15] The graphitic structure of the irradiated area (~500 $\mu m^2$) was confirmed by Raman scattering. Unfortunately, the spectra obtained at that time were not extended above 1800 $cm^{-1}$ and the typical graphene spectral features were not examined. From the above brief survey, it becomes clear that the use of focused laser beams, apart from enabling patterning, offer viable solutions to lower the graphene growth temperature down to ambient as well as to speed-up the growth process.

In the present article we demonstrate a novel facile method for the single-step, fast production of large area, homogeneous EG on the SiC(0001) using a continuous wave infrared $CO_2$ laser (10.6 $\mu m$) as a heating source. The process does not require high vacuum or strict sample-chamber conditions; it takes place under Ar gas flow at atmospheric pressure and temperature. The present method offers a number of advantages in relation not only to conventional EG growth methods but also compared with the UV-laser-assisted EG growth.[13] **(i)** Ambient growth of EG; UHV or high vacuum is not necessary. **(ii)** The $CO_2$ laser beam can induce EG growth within the second time scale, depending on laser power, over a **large area** (~3-4 mm$^2$). **(iii)** Irradiation takes place at one SiC face ($000\bar{1}$ in the present case), while graphene grows epitaxially at the opposite side (polished Si-face). As a result, the Si-face smoothness is preserved. This geometry (Scheme 1) enables the feasibility of writing graphene patterns on a SiC(0001) surface in contact with a substrate. **(iv)** The cooling rate, which is essential for the uniformity of the stresses that develop on EG, can be as high as ~600 K s$^{-1}$ The fast cooling rate might also affect the stacking order or stacking faults of Si-face EG, which is the dominant factor affecting carrier mobility.[8,11] **(v)** No pre-treatment (e.g. H$_2$ etching, and so on) of SiC surface is needed to obtain high quality graphene, in contrast to the vast majority of the previous studies.

The formation of few layer EG on SiC and its features were investigated by scanning electron microscopy (SEM), X-ray photoelectron spectroscopy (XPS), secondary ion-mass spectroscopy (SIMS), and Raman scattering.

## 2. Results and Discussion

Graphene was grown on SiC (0001) using the set-up shown in **Scheme 1**. The set-up is a modification of an experimental arrangement developed previously for melting refractive ceramics in controlled gas atmosphere at temperatures above 2000 K.[16] **Figure 1(a)** shows a representative micrograph of EG on 6H-SiC(0001). The image covers partly the surface of non-irradiated SiC (grey area), the irradiated area (1) and the border between the two regions (2). Actually, both the grey area (SiC) and the border region (2) correspond to the 6H-SiC(0001) surface which was not directly illuminated and was thus exposed to lower temperature due to the power density distribution of the heating $CO_2$ laser beam. At the non-irradiated area SiC undergoes milder annealing during the heating procedure, which causes purification from surface contamination and limited silicon desorption. As a result, wide, highly uniform, atomically flat terraces emerge.

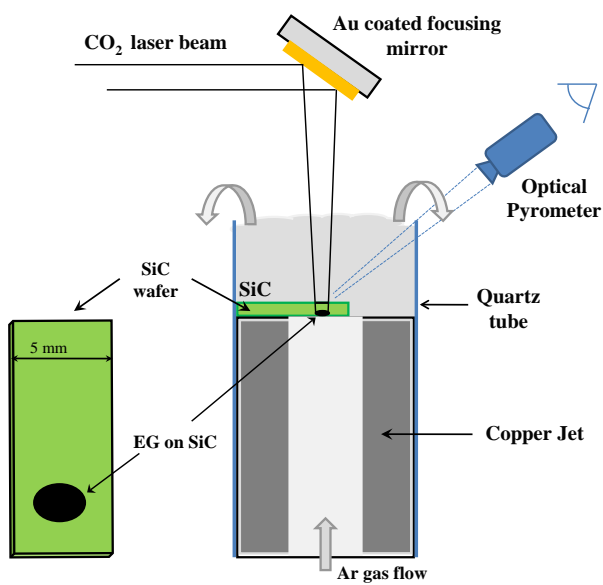

**Scheme 1:** Schematic diagram of the CO$_2$ laser induced epitaxial growth of graphene on SiC wafers.

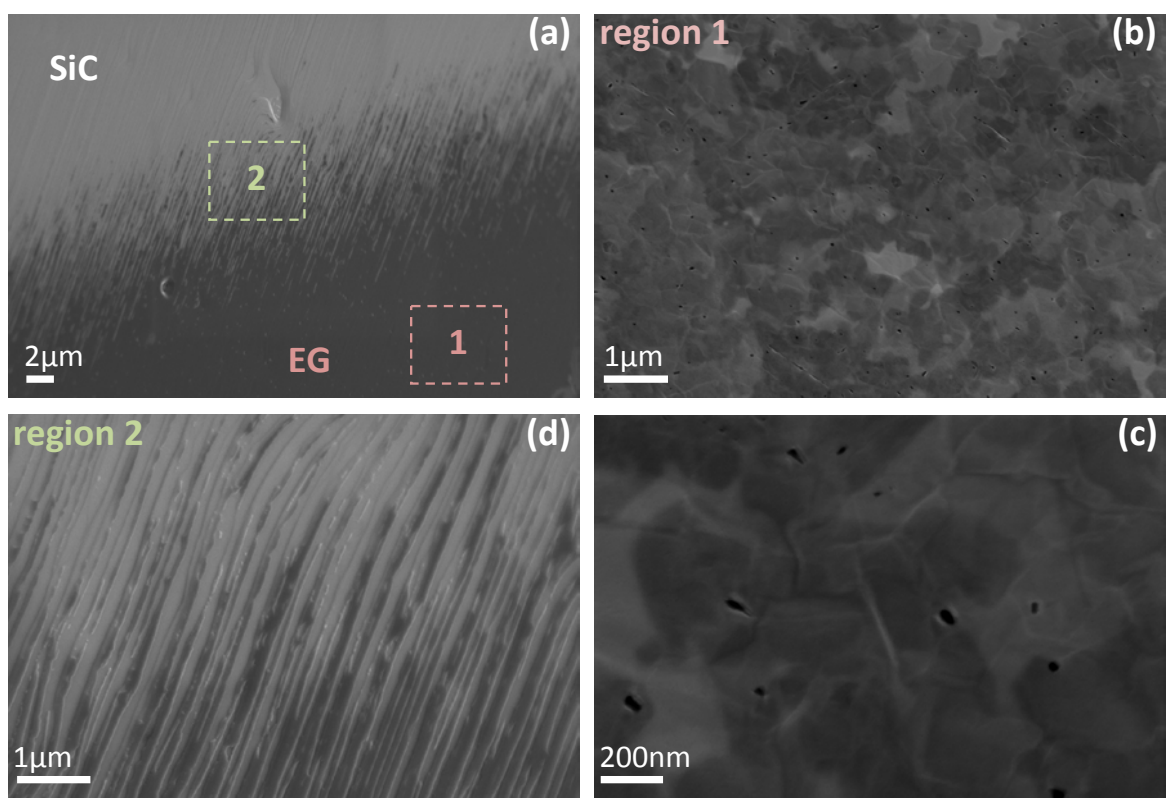

**Figure 1:** (a) SEM micrograph showing the formation of epitaxial graphene (dark area, 1) on 6H-SiC(0001) (grey area); (b), (c) detail from the irradiated area at high magnification; (d) close view of the borders (2) between the EG and SiC.

The first stages of graphene layer formation on these flat terraces are quite clear in Figure 1(d) which is an enlargement of the border area. Graphene grows initially at the terrace step edges where the large continuous terraces are found, extending parallel to the step edges. Graphene nucleation and growth is highly dependent on the SiC surface morphology. This type of growth has been previously observed for films grown under 900 mbar of argon on 6H-SiC(0001).[10]

Enlargement of the irradiated area is shown at low and high magnification in Figures 1(b) and 1(c), respectively. Similar characteristics to that observed on graphene films on SiC[17] and Ni[18] substrates are evident from these images. The terraces are not observable any longer, while the whole area is covered by the EG layers. The change of the image contrast from light to dark grey (or black) is directly correlated with the number of graphene layers on the substrate.[18] The lighter grey areas indicate thinner graphene film where the secondary electrons can more efficiently escape from the SiC substrate as compared with the darker ones.

Typical XPS C1s and Si2s spectra collected from the homogeneous large area of EG are shown in **Figure 2**. The Si2s photo-peak from the SiC substrate appears at binding energy 151.4 eV in agreement with literature values.[19] The spectrum does not contain components attributable to silica ($SiO_2$) or silicon oxycarbide ($SiC_xO_y$) which could be located at the interface between the SiC substrate and the EG overlayer. The C1s spectrum is analyzed into four components, at binding energies 283.5, 284.5, 285.5 and 287.2 eV originating from C atoms in the bulk SiC substrate, in the aromatic rings ($sp^2$ bonding) of the EG layer, in configurations with $sp^3$ bonding hybridization and in surface C=O species, respectively.[20] The peak at 285.5 eV has also been attributed to the first carbon layer grown on top of the SiC(0001) substrate forming strong covalent bonds with the latter and thus lacking the graphene electronic properties. This acts as a buffer layer, enabling the first graphene layer, to behave electronically like an isolated graphene sheet.[21]

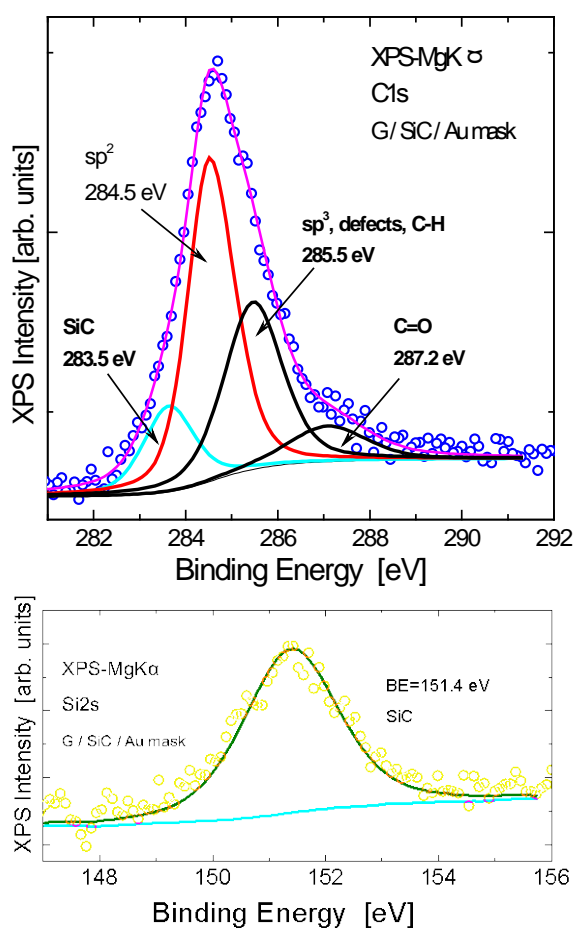

**Figure 2:** XPS core level spectra of Si2s (bottom) and C1s (top) of the EG/SiC sample (the non-irradiated area of SiC is covered with a gold mask).

Assuming that the graphene-SiC sample can be described as a semi-infinite SiC substrate with a uniform graphene overlayer of thickness $d$, the latter can be calculated from the total C1s peak intensities using the equation: $d = \lambda \ln[(I_{EG}C_{SiC}/I_{SiC}C_{EG})+1]$. $\lambda$ is the C1s electrons inelastic mean free path, assumed to be the same in SiC and in graphene; its values are reported elsewhere.[22] $I_{EG}$ and $I_{SiC}$ are the intensities (areas) of the C1s signal from the EG overlayer and the SiC substrate, respectively, while $C_{SiC}$ and $C_{EG}$ are the corresponding carbon atomic densities in each material.[23] The intensity ratio between the EG C1s and the SiC C1s lines was found to be $I_{EG} / I_{SiC} = 3.4$. For this ratio, the EG thickness is ~2.5±0.5 nm which, considering the thickness of a graphene layer to be equal to the interlayer distance in the graphitic structure (0.34 nm), corresponds to 5.5–8.5 monolayers. The relatively large error margin in this calculation emerges due to the uncertainty in the inelastic mean free path values and also because the estimate is based on a simple model of a bulk-terminated substrate. It is worth mentioning that the XPS signal is averaged over an area of ~2 mm² on which, the EG layer is considered to be homogenous.

Dynamic SIMS was used for depth profile measurements of the irradiated samples. **Figure 3** shows the depth distribution of amu 12, 41 and 28 recorded in the positive mode and representing $^{12}C^+$, $^{29}Si^{12}C^+$ and $^{28}Si^+$ or $^{12}C^{16}O^+$ species, respectively. The inset in Figure 3, is a chemical image of the sample. It shows five craters in the dark area (crater size ~158 μm). These are the spots where the depth repeated profile measurements took place. All spots produced similar profiles ascertaining the homogeneity of the irradiated area in depth. An additional depth profile figure from a different area of the sample is provided in the Supporting Information.

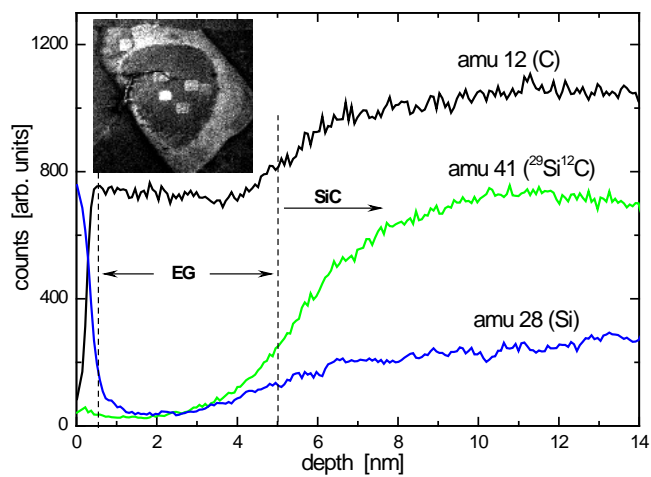

**Figure 3:** Dynamic SIMS depth profile analysis of the irradiated area of the SiC sample. Inset: Chemical image of the irradiated SiC sample. The grey colored area represents amu 16 ($^{16}O^-$) and the black domain shows amu $^{12}C^-$. The square craters are the sputtered spots on the sample.

The large signal of amu 28 in the initial steps of the procedure is attributed to surface species of oxidized carbon ($^{12}C^{16}O^+$) due to surface contamination.[24] It is observed that up to a depth of about 3 nm the signals of Si and SiC are close to the detection limit while that of carbon is considerably high and homogenous in depth. After the first 3 nm (i.e. ~9 ML of EG) the intensity of the Si-containing species starts to increase, indicating that the EG/SiC interface is reached. A certain plateau (bulk SiC) is attained after a depth of about 5.5-6 nm. When approaching the SiC substrate, the intensity of amu 12 increases as well, due to the higher sputter yield of carbon in the SiC matrix as compared to that in graphite.[25] From this profile the thickness of the EG layer appears to be in good agreement with the value estimated by XPS for the same sample. Details about SIMS measurements and depth calibration are given in the Supporting Information.

Raman spectra were recorded from several EG samples prepared in this work. Representative Raman spectra from various points of the SiC irradiated area are shown in **Figure 4**. For one of the samples the irradiated area of ~3 mm$^2$ was scanned systematically and Raman spectra measurements for over 100 different points were obtained (see Supporting Information). Data collected from the borders of the irradiated spot exhibit strong peaks arising from the second order Raman spectrum of SiC, on which the G-band of EG is superimposed (see Figure 4(a)). Comparing the relative intensities of the G peak and the SiC bands with corresponding spectra in previous studies our spectra could be interpreted as indicating the presence of 1 or 2 ML of EG at the borders of the illuminated area. For Raman spectra recorded from points located well within the spot, the SiC bands are much weaker than the graphene bands and hence are practically invisible; see lower spectrum in Figure 4(b). A line scan performed at the periphery of the irradiated area revealed rather sharp limits between these two types of Raman spectra. Practically, spectra as that shown in Figure 4(a) are recorded from a thin belt of about 10 μm in thickness around the irradiated area. This finding demonstrates homogeneous heating of the irradiated area with negligible lateral heat flow to the surrounding area.

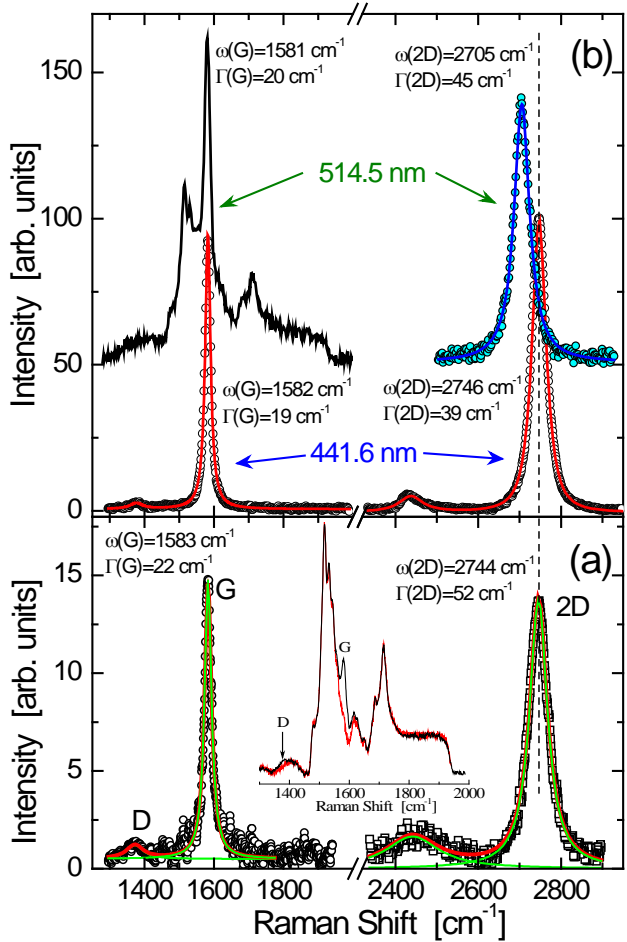

**Figure 4:** Representative Raman spectra of EG on SiC(0001) recorded (a) from the borders of the irradiated area and (b) from the irradiated spot. The upper spectrum in (b) was recorded with a macro-Raman set-up from a macroscopic sample area ~$10^5$ μm$^2$.

Extensive reviews on Raman scattering in MCG and EG have appeared over the last few years providing the means of characterizing the properties of graphene though the energy and width of certain Raman bands.[26-29] In brief, three main bands dominate Raman spectra of graphene: The G-band (~1582 cm$^{-1}$), present to all sp$^2$ carbon networks corresponds to the in-plane, doubly degenerate E$_{2g}$ phonon at the center of the Brillouin zone. The defect-induced D-band (dispersive mode in the range 1300-1370 cm$^{-1}$) arises from the breathing mode of the sp$^2$ rings and becomes Raman active in the presence of defects. The 2D band (the second order of D-band, sometimes referred to as the G´ band) originates from a double resonance process, being also dispersive; its presence in the Raman spectrum does not require the presence of the D band. A number of Raman studies have concentrated on EG,[30-36] which in combination with other techniques have provided information on graphene quality characterization by Raman scattering and phenomenological correlations between spectral Raman features and graphene properties. Combinations of the D, G, and 2D peak parameters have led to five main criteria which provide information on **(a)** disorder and crystallite size (D/G intensity ratio), **(b)** number of graphene ML (2D band shape and 2D/G ratio), **(c)** layer stacking order (2D band shape, i.e. single or multiple lines), **(d)** stresses (G band shift; shape and Raman shift of 2D band) and **(e)** carrier mobility (2D band width).[26-29] The analysis of our Raman spectra will proceed alongside the aforementioned phenomenological correlations/observations.

**(a)** In the vast majority of our Raman spectra, recorded well inside the irradiated area, (see Supporting Information) a very low D/G area ratio (~0.05) was found indicating rather large crystallite size $L_a$ and thus high quality graphene. Using the expression given by Cancado *et al*.[37] which relates crystallite size with D/G ratio we obtain $L_a \approx 182$ nm for the 441.6 nm excitation energy. This value indicates EG quality comparable with that reported elsewhere.[33,36] At the borders of the illuminated spot the D/G area ratio is of about 0.1, which is, however, much smaller than that reported in studies of furnace grown EG.[31] Furthermore, $CO_2$-laser-induced grown EG exhibits much lower disorder than UV-laser synthesized EG where the D/G area is of about unity.[13]

**(b)** As regards monolayer MCG there is consensus that the 2D/G intensity ratio assumes values near 4 (the ratio decreases with the number of MLs) and the 2D band shape is a single Lorentzian with full width at half maximum $\Gamma(2D) < 30$ cm$^{-1}$, which broadens appreciably so as four Lorentzians are needed for a correct fitting for bilayer MCG.[38] In general, the 2D/G ratio in EG is much lower than that in MCG; values around 0.5 or 2 are frequently reported.[33] However, this ratio is not considered as a valuable indicator of the number of MLs in EG. Our Raman spectra spanning the whole illuminated area show consistently 2D/G>1.5.

Controversial findings have been reported concerning the use of the 2D band-shape for estimating the number of MLs in EG. Faugeras et al.[32] studied EG grown on 4H-SiC($000\overline{1}$) reporting that the 2D peak is single Lorentzian, irrespective of the number of layers; the number of layers affects only its width. They reported $\Gamma(2D) \approx 60$ cm$^{-1}$ and $\approx 80$ cm$^{-1}$ for few-layer (up to 10) and multi-layer (90 layers) EG, respectively. Röhrl et al.[30] studied EG grown on 6H-SiC(0001) reporting that the 2D band is single Lorentzian for graphene ML, while it turns to a broader and asymmetric band composed of four Lorentzians for the bilayer and to even broader line for few (6) layer graphene. Single Lorentzian 2D band with $\Gamma(2D) \approx 60$ cm$^{-1}$ and 95 cm$^{-1}$ for ML and bilayer graphene grown on 6H-SiC(0001) was reported by Ni et al.[31] A more detailed investigation of monolayer and bilayer EG grown 6H-SiC(0001) and a tri-layer grown on 4H-SiC(0001) was undertaken by Lee et al.[33] They found that $\Gamma(2D) \approx 46$, 64, and 74 cm$^{-1}$ for the monolayer, bilayer and trilayer, respectively. The authors noted that up to 14 ML of EG, the 2D band does not develop any shoulder or asymmetry but it is poorly fitted by a single Lorentzian even for the bilayer which needs four Lorentzians for a satisfactory fitting. Lee et al.[33] found a correlation between the inverse number, $N$, of MLs and the width of the 2D band, i.e. $\Gamma(2D) = (-45N^{-1} + 88)$ cm$^{-1}$. The relation was derived for (Si-face) SiC(0001) EG while the data for SiC($000\overline{1}$) do not follow any systematic trend with N. A large number of Raman spectra recorded with the 441.6 nm excitation energy in this study showed that $\Gamma(2D) \approx 50$ cm$^{-1}$, although at few points $\Gamma(2D)$ was found narrower, i.e. of about 40 cm$^{-1}$. According to the above relation between $\Gamma(2D)$ and N the EG grown by the method of the current work would nominally consist of N$\approx$2 or lower, which is certainly lower than the XPS and SIMS result. This finding shows that spectroscopic techniques do not still provide an accurate determination of graphene MLs using spectroscopic techniques.

**(c)** Attempts to study the number of ML in graphene have up to now focused on samples with AB Bernal stacking, and this is mainly the case of MCG. When graphene layers are randomly rotated (turbostratic graphene) a single Lorentzian line can adequately fit the 2D band due to the decreasing inter-layer interactions. Similarities between EG on SiC($000\bar{1}$) and turbostratic graphite were noticed in view of the single Lorentzian, wide 2D band.[25] It has also been stated that EG is characterized by disordered stacking only when grown on C-face and not on the Si-face SiC(0001).[1,8] UV-laser grown EG was found to exhibit large $\Gamma$(2D), i.e. ~60 cm$^{-1}$ for the ML and ~75 cm$^{-1}$ for 2-3 ML.[13] Our results, i.e. the fact that the 2D band for all Raman spectra is a single Lorentzian line, demonstrate that few-layer EG with disordered stacking or stacking faults can also be produced on SiC(0001). The origin of this effect is still speculative; however, we suggest that the extremely faster heating and cooling rates offered by the method employed here, in comparison with the furnace treated graphenes, might provide a rational explanation for the emergence of non-Bernal stacking. This stems as an important result given that the existence of rotational stacking faults has been shown to provide EG grown on SiC electronic properties similar to those of the isolated single graphene layer.[39, 40]

**(d)** A great deal of controversy can be found also in the literature regarding the use of the frequency of the G- and 2D-band and its shift in relation to that of MCG monolayers, in the determination of stresses in EG grown on SiC. While Faugeras *et al.*[32] reported that the G-band energy (1582 cm$^{-1}$) is the same for MCG, graphite and EG grown on SiC($000\bar{1}$), other works for EG on 6H-SiC(0001) showed that $\omega$(G) can be higher, i.e. $\omega$(G) $\approx$ 1608,[30] 1597,[31] and 1591[33] cm$^{-1}$. According to Rohrl *et al.*[30] $\omega$(G) red-shifts with increasing number of graphene layers tending to the graphite value. As regards the 2D band, contradictory results appear also: $\omega$(2D) was found to be almost fixed in the interval 2655–2665 cm$^{-1}$ (for 632.8 nm excitation);[32] $\omega$(2D) decreases with increasing the number of EG layers;[30] $\omega$(2D) increases by ~40 cm$^{-1}$ from the monolayer to the trilayer;[33] and ω(2D) increases by ~21 cm$^{-1}$ from the monolayer to the bilayer, both being lower than the ω(2D) of MCG.[31] Compressive strain and in few cases electron/hole doping have been considered in most of the above studies to account for the differences of ω(G) and ω(2D) between EG and MCG.

In our measurements we find that ω(G) ≈ 1582±1 cm$^{-1}$ and ω(2D) ≈ 2745±3 cm$^{-1}$ (for 441.6 nm laser energy) for all Raman spectra recorded from a large area of EG. This finding points to a very uniform distribution of strain along the whole area where EG has grown. To confirm this we present in Figure 4(b) the Raman spectrum recorded with the aid of a macro-Raman set-up using the excitation line 514.5 nm. This is to our knowledge the first Raman spectrum from a really macroscopic EG scattering area (10$^5$ μm$^2$). The sample was scanned at few various, not overlapping, points exhibiting indistinguishable Raman spectra. The 2D band is red-shifted, relative to the value obtained with the 441.6 nm laser line, owing to its dispersive nature to ω(2D) ≈ 2705 cm$^{-1}$, while its width is similar, Γ(2D) ≈ 45 cm$^{-1}$, to the other values recorded from a limited spatial region. The most striking result is that its band-shape can be excellently fitted by a single Lorentzian. These findings demonstrate that the present EG preparation method yields very homogeneous EG as regards the developed strain. Finally, it is instructive to note that a large shift of ω(2D) of about 30 cm$^{-1}$ for 1 ML and 2-3 ML EG is reported for UV-laser synthesized EG.[13]

In a detailed spatially resolved Raman study[34] of EG grown on 6H-SiC(0001) it was found that EG strain varies substantially between various points few μm apart, i.e. the 2D frequency can vary by ~65 cm$^{-1}$. Even for points differing for 600 nm the strain was significant to cause shifts of the 2D band by ~25 cm$^{-1}$. This shows that furnace produced EG can be highly non-uniform at the μm scale. Robinson *et al*.[34] categorized EG as high strain or low strain when the 2D band energy is higher or lower, respectively, than the corresponding band in bulk graphite; while EG would be strain free when ω(2D) approaches that of MCG.

Their study showed that high and low strain areas coexist at comparable fractions. Considering that (at 514.5 nm excitation) $\omega(2D) \approx 2687$ cm$^{-1}$ for free standing monolayers of MCG,[28] and $\omega(2D) \approx 2726$ cm$^{-1}$ for bulk graphite,[28] it can be concluded that our $\omega(2D) \approx 2705$ cm$^{-1}$ supports that the $CO_2$-grown EG on SiC(0001), apart from being uniform at the macro-scale, experiences also very low-strain. Finally, compressive strain which is usually adopted to account for EG Raman band shifts, would affect (shift) simultaneously the G and 2D bands with a shift ratio of about 1/2.5.[41] The absence of shift for the G-band renders the hypothesis of strain in the $CO_2$-laser synthesized EG questionable. Apart from compressive strain, charge doping from the substrate, and/or doping emerging from the oxygen/ water from the atmosphere, can account for the blue-shift of the 2D band. The finding by Lee *et al.*[33] that $\omega(2D)$ does not red-shifts substantially after transferring EG from SiC to SiO$_2$ substrates led to the conclusion that SiC does not exert any appreciable strain on EG. Therefore, we cannot exclude the possibility that the small shift (18 cm$^{-1}$) of the 2D band, between EG and MCG (for 514.4 nm) could originate from doping (carriers provided by SiC).

(e) A remarkable correlation was proposed between the carrier mobility and $\Gamma(2D)$ in EG grown on SiC($000\overline{1}$).[35] $\Gamma(2D)$ was considered to be affected by the stacking order of graphene layers. Narrow single Lorentzian 2D bands ($\Gamma(2D) < 40$ cm$^{-1}$) originate from rotationally faulted multilayer EG. Broad, asymmetric 2D-band shapes were interpreted as originating from the simultaneous contribution of both rotationally faulted and Bernal stacked. Comparing $\Gamma(2D)$ and carrier mobility data, it was found that sample regions with narrow $\Gamma(2D)$ exhibit high carrier mobilities, while in samples areas where Bernal stacking also contributes to $\Gamma(2D)$ the mobility is severely lower.[35] The correlation holds only for uniform sample areas exhibiting less than 10% variation in $\Gamma(2D)$. Our findings show that EG grown on SiC(0001) by $CO_2$ laser heating can be considered as uniform based on the negligible (much less than 10%) spatial variation of $\Gamma(2D)$ over large areas. However, although our 2D bands can be satisfactorily fitted by a single Lorentzian, the width is higher than the corresponding bands in Ref. [35] in the regime where rotational faults dominate. It would thus be unsafe to draw any conclusion of carrier mobility in the EG grown by the current method simply based on $\Gamma$(2D), especially considering that areas in the scattering volume which contain EG grown on the edges of the terraces (with different curvature) can also contribute to the 2D-band broadening.[33]

## 3. Conclusions

In summary, we presented a novel method for the fast, one-step epitaxial growth of large-area homogeneous graphene on SiC(0001). The use of a $CO_2$ laser as the heating source provides flexibility in growing uniform, low–strain graphene film composed of few monolayers. The method is quite simple and cost-effective in that it does not necessitate SiC pre-treatment and/or high vacuum, it operates at low temperature and proceeds in the second time scale, thus providing a green solution to EG fabrication and a means to engineering graphene patterns on SiC by focused laser beams. The very high heating rate achieved by $CO_2$-laser heating seems to have beneficial results in avoiding the different Si desorption rates from adjacent SiC steps,[42] while the corresponding ultrafast cooling rate influences the stacking order of EG on SiC(0001). Work underway is focused on controlling EG thickness by changing illumination time and/or laser power density.

## 4. Experimental

*4.1 Epitaxial Graphene synthesis on 6H-SiC(0001)*: Graphene was grown on 260 μm thick *n*-doped 6H-SiC wafers, Si-face polished with orientation 0°, 20´ with respect to the ideal (0001) plane. Rectangular strips (~5×10 mm$^2$) cut from the wafer, were washed with HF solution and rinsed with ethanol. The SiC strip was placed on top of an opening of a small copper metal support where Ar gas was flowing through, thus creating an inert gas blanket around the sample at atmospheric pressure. The laser beam was focused with a gold coated concave mirror on the C-terminated ($000\bar{1}$) face (unpolished wafer side) of the SiC strip producing a visible elliptical trace with short and long axes of ~1.5 and ~2.5 mm, respectively (Figure S1). The temperature of the spot was monitored with a two-color optical pyrometer (model OS 3722, Omega Engineering Inc.). In a series of experiments different percentages of the $CO_2$ laser power were utilized and the sample was irradiated at different time intervals reaching temperatures up to ~2400 K. The modification of the SiC surfaces became (visually) detectable at temperature above ~2000 K and at exposure times of few seconds. However, it was found by various experiments that a rather uniform modification of the SiC surfaces occurs when about 40% of the maximum $CO_2$ laser power was utilized and the irradiation lasted for ~10 s. The laser power was preset at this level with the beam blocked and the illumination of the C-face of the SiC wafer was sudden. Under these conditions, the illuminated volume reaches the maximum temperature of ~2400 K with a heating rate of ~600 K s$^{-1}$. At the end of the heating procedure the illuminated SiC surface became slightly blackish as shown in Figure S1. The spot on the reverse surface (Si-face) of the SiC strip was rather uniform and smooth and was the subject of our studies. Raman scattering and XPS showed no evidence of graphitic material on the C-face of the SiC wafer. This can be attributed to the fact that the atmospheric $O_2$ partial pressure is significantly higher on the C-face of SiC – which is directly exposed to the $CO_2$ laser beam – in comparison to that of the Si-face.

*4.2 EG characterization by SEM, XPS and SIMS*: Scanning electron microscopy was carried out with a high resolution field emission SEM (Zeiss SUPRA 35VP) operated at 15 kV. XPS measurements were carried out in an ultrahigh vacuum chamber equipped with a SPECS LHS-10 hemi-spherical electron analyzer and a dual x-ray anode, using the unmonchromatized MgK$_\alpha$ (1253.6 eV) line. The analyzer operated at the constant pass energy mode, $E_p$ = 36 eV giving a full width at half maximum of 1.1 eV for the main C1s XPS peak. The XPS core level spectra were analyzed using a fitting routine, which can decompose each spectrum into individual mixed Gaussian-Lorentzian peaks after a Shirley background subtraction. Regarding the measurement errors, for the XPS core level peaks we estimate that errors in peak positions are of about ±0.05 eV for a good signal to noise ratio. For the photoemission measurements the sample was wrapped by a gold foil mask covering all the untreated SiC substrate and having an opening with diameter ~1.5 mm to expose about 60% of the treated sample area. A drawing of the sample is shown in Scheme S1. Dynamic SIMS were performed at the Millbrook MiniSIMS spectrometer using a liquid gallium ion gun ($E$=6 keV). The mass analyzer was a quadruple filter with mass range of 2–300 Da.

*4.3 Raman Spectroscopy*: The overall area where the EG growth took place was monitored by micro-Raman spectroscopy. The sample was mounted on the calibrated XY stage of the microscope and consecutive measurements were performed at selected points of the square grid superimposed on the image shown in Figure S1. Over a hundred of Raman spectra of EG were systematically measured from different points from the sample shown. Raman measurements were repeated in few other samples prepared in a similar way. The majority of the spectra were collected with the Labram HR800 (Jobin-Yvon) micro-Raman system equipped with a He-Cd laser operating at 441.6 nm. Circumstantially, Raman spectra were recorded with the 514.5 nm laser line using the T64000 (Jobin-Yvon) Raman spectrometer in the micro- and macro-configurations. Microscope objectives with magnifications 100× and 50× were used to focus the light onto a ~1 and ~3 μm spots, respectively. In the macro-Raman measurements, the light was incident on the EG surface at an angle of ~60° in relation to the normal of the SiC wafer and the scattered light was collected at right angle. The thus formed laser spot area on the EG surface was of about $10^5$ μm$^2$ covering a large fraction of the sample area. The large penetration depth of the laser line at this configuration is responsible for the SiC modes appearing around the G-band spectral range. Furthermore, in selected areas 5 to 10 consecutive spectra were recorded every ~5-10 μm on a line. The uniformity (homomorphism) – as regards the energy and the shape of the G and 2D bands – of the measured spectra for selected points on the grid is evident from Figure S3. The figure contains representative spectra – out of the over 100 spectra measured – along four lines, A, B, C, and D as marked in Figure S1. For each line, the three spectra shown were recorded from areas near the borders and around middle of the illuminated spot. The second order Raman bands of SiC is evident in all spectra recorded from the edges of the EG spot. Representative Raman spectra recorded with the 514.5 nm laser are shown in Figure S4, where the first order Raman spectrum of SiC is shown at wavenumbers below 1000 cm$^{-1}$.

# SUPPORTING INFORMATION for

## $CO_2$ Laser-Induced Growth of Epitaxial Graphene on SiC(0001)

*Spyros N. Yannopoulos, Angeliki Siokou, Nektarios K. Nasikas, Vassilios Dracopoulos, Fotini Ravani and George N. Papatheodorou*

### 1. 6H-SiC(0001) absorptance at 10.6 μm

The optical properties of SiC are crucial for the efficient and fast heating of SiC via the absorption of the $CO_2$ laser infrared radiation. The reflectance (R) spectrum of SiC, taken from (SR1) and shown in Figure S2, exhibits a very sharp profile between 10 and 11 μm. The reflectance of 6H-SiC is ~90% at the laser wavelength, i.e. 10.6 μm. With only 10% of the incident radiation absorbed, it seems unlikely that the use of a $CO_2$ laser line could efficiently raise the temperature. However, Roy *et al*. (SR2) have shown that the reflectance is significantly blue-shifted with increasing temperature (R≈0.15 at 1273 K). This causes more efficient absorption and heating of SiC leading to its decomposition.

Pre-treatment of the SiC surface, in order to remove polishing damage, is considered as an important step before the growth of EG. Previous studies have shown that EG grown on SiC(0001) surfaces without $H_2$ etching exhibit significant degradation of the SiC surface during graphitization leading to the formation of highly non-uniform patterns (SR3). However, in the present study, SEM images confirm that even without this time consuming procedure, good quality EG can grow on SiC(0001).

### 2. EG thickness determination by XPS and SIMS

The thickness of the EG film was evaluated by XPS and dynamic SIMS measurements. As mentioned in the main text, the C1s spectrum is analyzed into four components, at binding energies 283.5, 284.5, 285.5 and 287.2 eV originating from C atoms in the bulk SiC substrate, in the aromatic rings ($sp^2$ bonding hybridization) of the EG layer, in configurations with $sp^3$ bonding hybridization and in surface C=O species, respectively. In order to evaluate the credibility of the C1s peak analysis using an independent parameter, we attempt to calculate the Si/C atomic ratio in the bulk substrate by the intensity ratio, $I_{Si2s}/I_{C1s(SiC)} = 0.75$, between the Si2s XPS peak (bottom panel in Figure 2 in the main text) and the SiC component of the C1s spectrum. Assuming a homogenous graphitic layer with thickness of 2.5 nm (the calculation of this thickness is presented in the main text) on top of the bulk SiC and using the relative sensitivity factors for Si2s and C1s from Ref. (SR4) corrected for the SiC matrix, and with λ values taken from Ref. (SR5) we calculate the atomic ratio in the bulk silicon carbide substrate to be Si/C=1.0±0.05. The agreement of this value with the nominal stoichiometry of the substrate affirms the credibility of the C1s peak analysis and consequently of the C1s intensity values used for the calculation of the graphene layer thickness.

Additional assessment of the EG layer thickness was performed by dynamic SIMS using the Millbrook MiniSIMS spectrometer. MiniSIMS raw depth profiling data (see Figure 3 in the text and Fig. S3) are recorded in units of counts per second (cps) versus etch-time (or sputtered time). The conversion of the etch-time to depth is done by calculating a sputter rate which is *matrix dependent*. Consequently for the calibration of the depth axis in the present case, two different sputter rates for carbon are required. The one within the SiC matrix was calculated by measuring the depth – using a profilometer (WYKO NT 1100, Veeco Inc.) – of the crater formed on a non-irradiated 6H-SiC(0001) crystal which was sputtered at the same conditions as the irradiated sample. In order to estimate the sputter rate of carbon from the EG matrix, the same procedure was attempted using a model HOPG surface. Nevertheless, the craters formed on HOPG appeared to have uneven depth, making the accurate evaluation of their size tricky and inaccurate. Nevertheless, according to the literature the sputter yield of carbon from a SiC matrix is higher than that from graphite (SR6) indicating that the thickness found for the EG layer from these measurements is slightly overestimated.

### 3. Raman measurements

Figure S4 show typical Stokes-side micro-Raman spectra of EG on 6H-SiC(0001) at various points of the grid denoted in Fig. S1 recorded with the 441.6 nm laser line at backscattering geometry. Representative Raman spectra recorded with the 514.5 nm laser are shown in Fig. S5, where the first order Raman spectrum of SiC is shown at wavenumbers below 1000 cm$^{-1}$.

**Supplementary Figure Captions**

**Supplementary Figure S1.** Optical image of the 6H-SiC(0001) wafer surface on which EG was grown by $CO_2$ irradiation. The dark elliptical spot between the horizontal borders A, F and the vertical borders 2, 10 represents the area (~3 mm$^2$) where EG has grown. The superimposed grid has a node-to-node distance of 300 μm.

**Supplementary Figure S2.** Reflectance spectrum of SiC taken from (SR1). The vertical line indicates the wavelength of the $CO_2$ laser line.

**Supplementary Scheme S1:** Drawing of the SiC sample with the EG spit in the middle, covered with the Au mask for the XPS experiment.

**Supplementary Figure S3.** Dynamic SIMS depth profile analysis of the irradiated area of the SiC sample.

**Supplementary Figure S4.** Representative Stokes-side micro-Raman spectra (441.6 nm) of EG on 6H-SiC(0001) spanning the whole illuminated area. The dashed vertical lines in the middle column denote the fixed energies of the G and 2D bands. Labels in each panel correspond to the nodes of Fig. S1, from where each spectrum was recorded.

**Supplementary Figure S5.** Micro-Raman spectra of EG excited with the 514.5 nm laser line. (a) Spectra at the border of the illuminated area where the EG spectrum overlaps with that of the SiC substrate. (b) Spectrum inside the illuminated area.

**Figure S1**

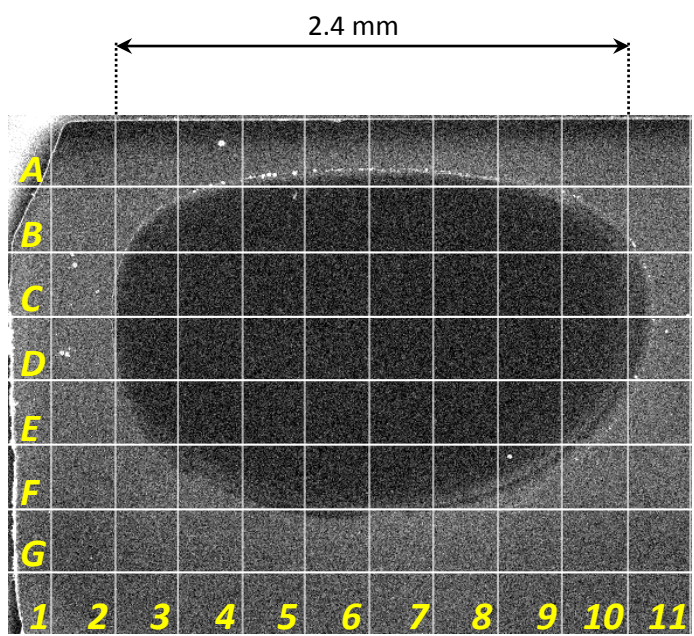

**Figure S2**

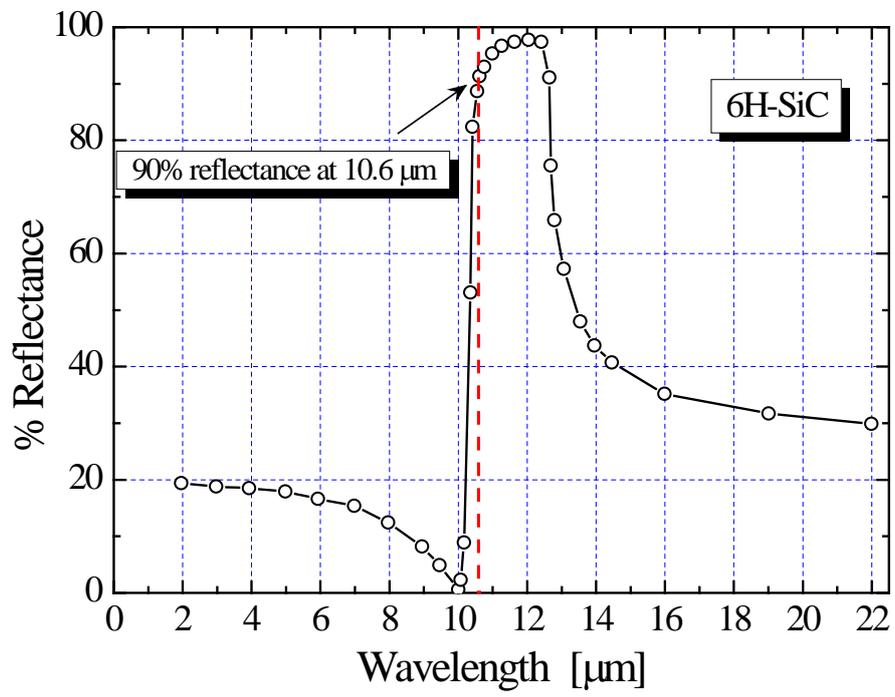

**Scheme S1**

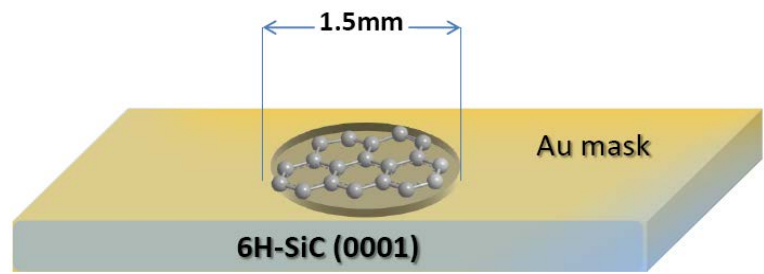

**Figure S3**

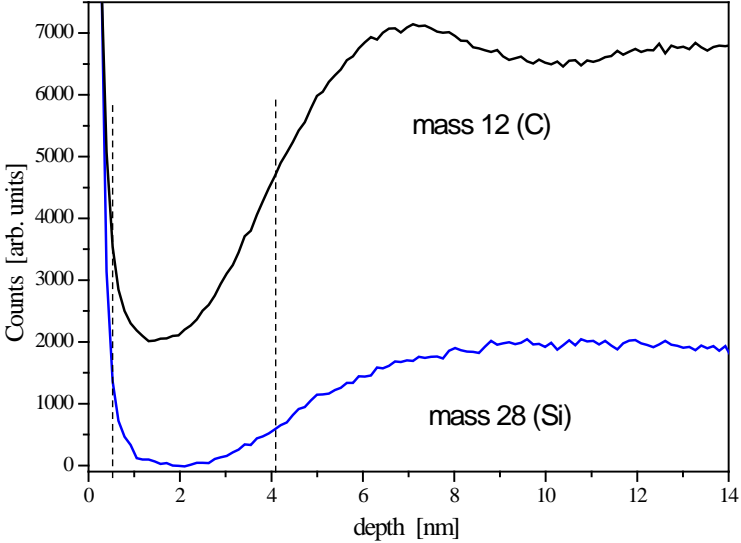

**Figure S4**

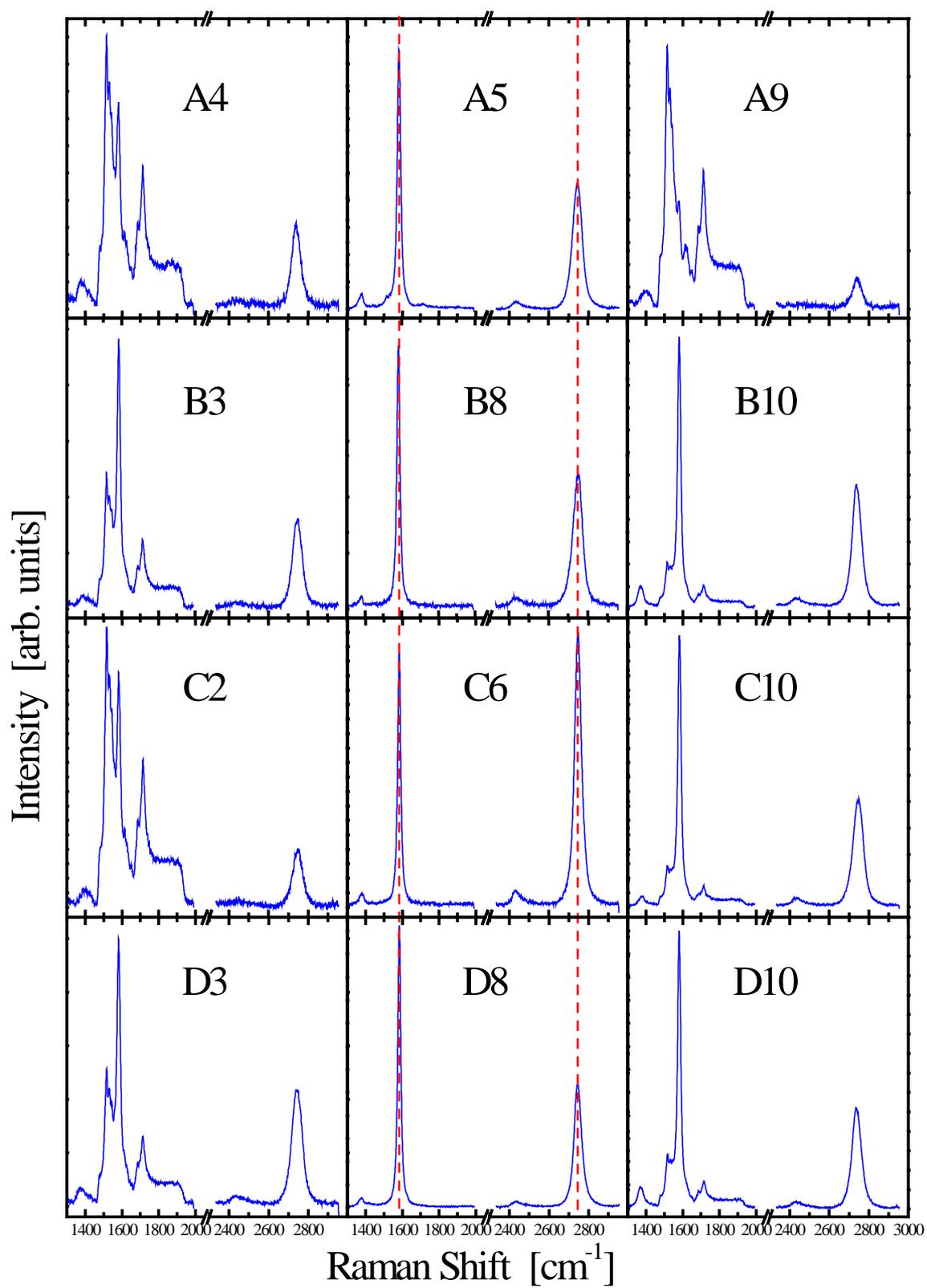

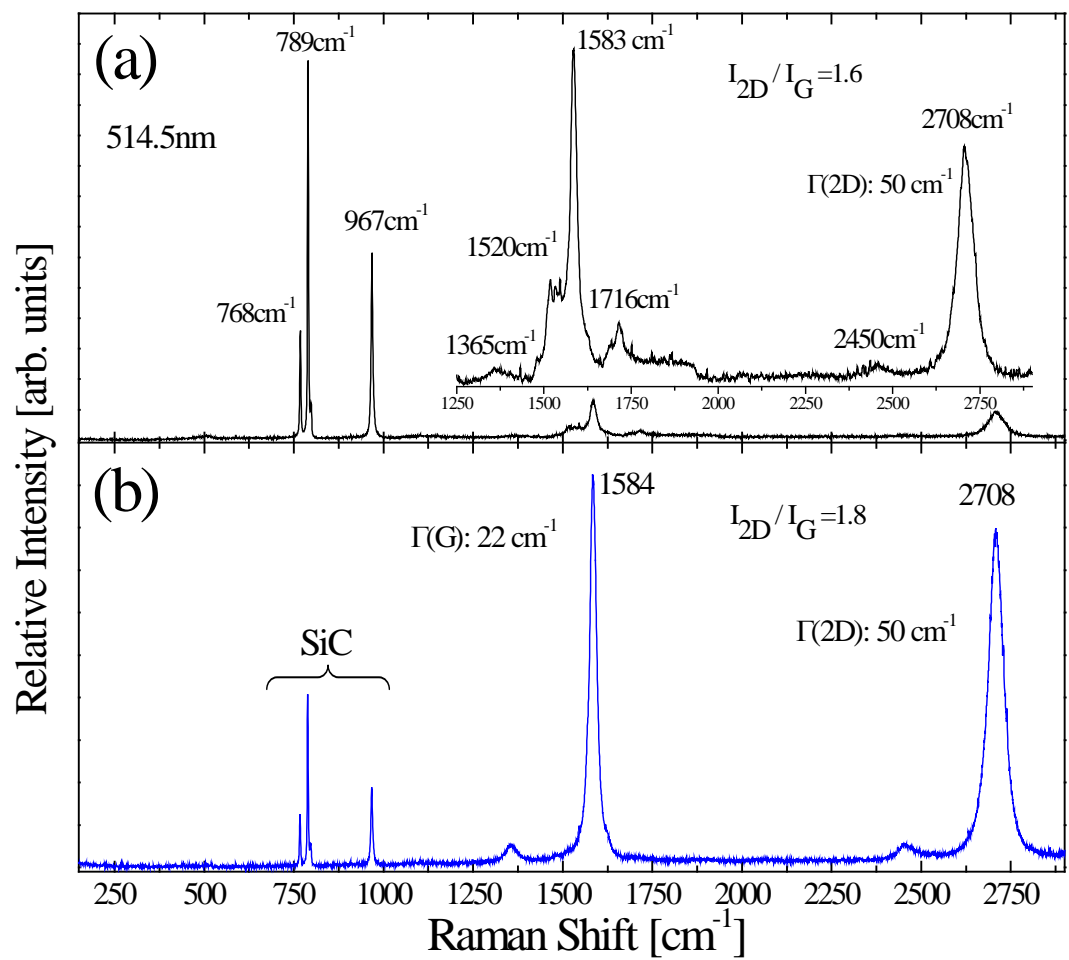